\documentclass[prd,preprint,showpacs,amssymb]{revtex4}
\usepackage{graphicx}
\begin{document}

\title{Magnetic Phases in Three-Flavor Color Superconductivity}
\author{Efrain J. Ferrer}
\author{Vivian de la Incera}
\affiliation{Department of Physics, Western Illinois University,
Macomb, IL 61455, USA}

\begin{abstract}
The best natural candidates for the realization of color
superconductivity are quark stars -not yet confirmed by observation-
and the extremely dense cores of compact stars, many of which have
very large magnetic fields. To reliably predict astrophysical
signatures of color superconductivity, a better understanding of the
role of the star's magnetic field in the color superconducting phase
that realizes in the core is required. This paper is an initial step
in that direction. The field scales at which the different magnetic
phases of a color superconductor with three quark flavors can be
realized are investigated. Coming from weak to strong fields, the
system undergoes first a symmetry transmutation from a
Color-Flavor-Locked (CFL) phase to a Magnetic-CFL (MCFL) phase, and
then a phase transition from the MCFL phase to the Paramagnetic-CFL
(PCFL) phase. The low-energy effective theory for the excitations of
the diquark condensate in the presence of a magnetic field is
derived using a covariant representation that takes into account all
the Lorentz structures contributing at low energy. The field-induced
masses of the charged mesons and the threshold field at which the
CFL $\rightarrow$ MCFL symmetry transmutation occurs are obtained in
the framework of this low-energy effective theory. The relevance of
the different magnetic phases for the physics of compact stars is
discussed.

\end{abstract}
\maketitle

\section{Introduction}
At present the physics community is actively trying to find ways to
differentiate a neutron star made up entirely of nuclear matter from
one with a color superconducting \cite{CS} quark matter core.
Compact stars typically have very large magnetic fields. Hence any
predicted signature of a color superconducting core should take into
account the presence of the star's magnetic field and its effects in
the superconducting state. Given that magnetars can have surface
fields as large as $10^{14} - 10^{16}$ G \cite{magnetars}, it is
reasonable to expect that the star's interior fields can reach even
higher values. Maximum strengths of $10^{18}-10^{19}$ G are allowed
by a simple application of the virial theorem \cite{virial}.

Although a color superconductor (CS) is in principle an electric
superconductor, because the diquark condensate carries nonzero
electric charge, in the color-flavor-locked (CFL) phase
\cite{alf-raj-wil-99/537} (CFL is the color superconducting phase
that realizes in a system of three-flavor massless quarks at high
densities) there is no Meissner effect for a new in-medium
electromagnetic field $\widetilde{A}_{\mu}$. This in-medium
electromagnetic field -called in the literature a "rotated"
electromagnetic field, where the "rotation" takes place in an inner
space- is a combination of the regular electromagnetic field and the
$8^{th}$ gluon \cite{alf-raj-wil-99/537,Gorbar-2000}. As the quark
pairs are all neutral with respect to the "rotated" electromagnetic
charge $\widetilde{Q}$, the "rotated" electromagnetic field
$\widetilde{A}_{\mu}$ remains long-range within the superconductor.

In this paper we are interested in the color-superconducting
magnetic phases that realize in a very dense system of three-flavor
massless quarks interacting in the background of a rotated magnetic
field $\widetilde{B}$. As shown in Ref \cite{MCFL}, the
color-superconducting properties of such a system are substantially
affected by the penetrating $\widetilde{B}$ field and as a
consequence, a new phase, called Magnetic Color Flavor Locked (MCFL)
phase \cite{MCFL}, takes place. In the MCFL phase the pairing of
(rotated) electrically charged quarks is reinforced by the field.
Pairs of this kind have bounding energies which depend on the
magnetic-field strength and are bigger than the ones existing at
zero field. At field strengths of the order of the baryon chemical
potential, the pairing reinforcement is sufficient to produce a
distinguishable splitting of the gap in two pieces: one that only
gets contributions from pairs of neutral quarks and one that gets
contributions from both pairs of neutral and pairs of charged
quarks.

Although the symmetry breaking patterns of the MCFL and CFL phases
are different, the two phases are hardly distinguishable at weak
magnetic fields. In the CFL phase the symmetry breaking is given
by
\begin{equation}\label{CFL}
\mathcal{G}=SU(3)_C \times SU(3)_L \times SU(3)_R \times U(1)_B
\times U(1)_{\rm e.m.}\rightarrow SU(3)_{C+L+R}\times {\widetilde
U(1)}_{\rm e.m.}.
\end{equation}
This symmetry reduction leaves nine Goldstone bosons: a singlet
associated to the breaking of the baryonic symmetry $U(1)_B$, and
an octet associated to the axial $SU(3)_A$ group.

Once a magnetic field is switched on, the difference between the
electric charge of the $u$ quark and that of the $d$ and $s$
quarks reduces the original flavor symmetry of the theory and
consequently also the symmetry group remaining after the diquark
condensate is formed. Then, the breaking pattern for the
MCFL-phase \cite{MCFL} becomes
\begin{equation}\label{MCFL}
\mathcal{G_{B}}=SU(3)_C \times SU(2)_L \times SU(2)_R \times
U(1)^{(1)}_A\times U(1)_B \times U(1)_{\rm e.m.} \rightarrow
SU(2)_{C+L+R} \times {\widetilde U(1)}_{\rm e.m.}.
\end{equation}
The group $U(1)^{(1)}_A$ (not to be confused with the usual
anomaly $U(1)_{A}$) is related to the current which is an
anomaly-free linear combination of $s$, $d$, and $u$ axial
currents \cite{miransky-shovkovy-02}. In this case only five
Goldstone bosons remain. Three of them correspond to the breaking
of $SU(2)_A$, one to the breaking of $U(1)^{(1)}_A$, and one to
the breaking of $U(1)_B$. Thus, an applied magnetic field reduces
the number of Goldstone bosons in the superconducting phase, from
nine to five.

The MCFL phase is not just characterized by a smaller number of
Goldstone fields, but by the fact that all these bosons are
neutral with respect to the rotated electric charge. Hence, no
charged low-energy excitation can be produced in the MCFL phase.
This effect can be relevant for the low energy physics of a color
superconducting star's core and hence for its transport
properties. In particular, the cooling of a compact star is
determined by the particles with the lowest energy; so a star with
a core of quark matter and sufficiently large magnetic field can
have a distinctive cooling process.

More recently, we have found that the magnetic field can also
influence the gluon dynamics \cite{Vortex}. At field strengths
comparable to the charged gluon Meissner mass a new phase can be
realized giving rise to an inhomogeneous condensate of
$\widetilde{Q}$-charged gluons \cite{Vortex}. The gluon condensate
anti-screens the magnetic field due to the anomalous magnetic
moment of these spin-1 particles. Because of the anti-screening,
this condensate does not give a mass to the $\widetilde{Q}$
photon, but instead amplifies the applied rotated magnetic field.
This means that at such applied fields the CS behaves as a
paramagnet, thus we named this phase paramagnetic CFL (PCFL)
\cite{PCFL}. This last effect is also of interest for
astrophysics. Compact stars with color superconducting cores could
have larger magnetic fields than neutron stars made up entirely of
nuclear matter, thanks to the gluon vortex antiscreening
mechanism.

The above state of affairs underlines the need to discern the
scales and field strengths at which one or another magnetic phase
is physically relevant. If one ignores the quark masses, the main
scales of the color superconductor are the baryon chemical
potential $\mu$, the dynamically generated gluon mass $m_{g}\sim
g\mu$ and the gap parameter $\Delta\sim \frac{\mu}{g^{5}}
e^{-\alpha/g}$, with $\alpha$ a constant that is dominated by
magnetic gluon exchanges \cite{Son}. We can assume that at
sufficiently high $\mu$, the running strong coupling $g$ becomes
$g(\mu)\ll 1$, so the the hierarchy of the scales is $\Delta\ll
m_{g}\ll \mu$.  The main purpose of this paper is to elucidate how
the different magnetic phases are related to the fundamental
scales of the CS.

The plan of the paper is as follows. In Sec. \ref{low-energy theory}
we develop the CFL low-energy effective theory in a magnetic
background using a Lorentz covariant formalism. In this derivation
the rotated magnetic field is introduced only through covariant
derivatives, thus preserving the $\widetilde{U}(1)_{em}$ gauge
invariance. The threshold rotated magnetic field that decouples the
charged mesons from the low-energy theory is found in Sec.
\ref{threshold-field}. In Sec. \ref{mag-phases} we discuss how the
different magnetic phases are realized in a hierarchical order
determined by the main energy scales of the CS. In the concluding
remarks we state the major outcomes of the paper and discuss
possible astrophysical implications of the realization of a
PCFL-like phase at moderate densities.

\section{Low-Energy Effective Theory in a Magnetic
Background}\label{low-energy theory}

The physics at energies below the lower scale $\Delta$ can be
explored by constructing the effective low-energy theory in the
presence of a rotated magnetic field. Since in the CFL phase all
the fermions are gapped and all the gluons have dynamically
generated masses, the low energy theory of the CS is governed by
the Goldstone modes arising from the breaking of the global
symmetries in the presence of a magnetic field.

As discussed in the Introduction, once a magnetic field is
present, the original symmetry group $\mathcal{G}$ is reduced, due
to the different electric charges of the quarks, to
$\mathcal{G_{B}}$. One would think that the low energy theory
should correspond to the breaking pattern (\ref{MCFL}), hence be
described by five neutral Goldstone bosons. However, it is clear
that at very weak magnetic fields the symmetry of the CFL phase
can be treated as a good approximated symmetry, meaning that at
weak fields the low-energy excitations are essentially governed by
nine approximately massless scalars (those of the breaking pattern
(\ref{CFL})) instead of five.

A question of order here is: what do we exactly understand as a very
weak magnetic field? In other words, what is the threshold-field
strength that effectively separates the CFL low energy behavior from
the MCFL one? A fundamental clue in this direction will come from
the determination of the term in the low-energy CFL Lagrangian that
can generate a field-induced mass for the charged Goldstone fields,
disconnecting them from the low-energy dynamics at some field
strength and thereby effectively reducing the number of Goldstone
bosons from the nine of the CFL phase, to the five of the MCFL. A
similar approach was previously followed in Ref. \cite {Manuel}.
Nevertheless, as it will become clear below, our treatment and
results will differ from those previously obtained.

Our strategy will consist of writing the effective low energy
Lagrangian for the Goldstone bosons corresponding to the CFL
breaking pattern (\ref{CFL}), but in the presence of an external
$\widetilde{B}$ field, that is, ignoring the explicit breaking
introduced by the electromagnetic interaction. To ensure that this
Lagrangian is "invariant" with respect to the original group of the
CFL case, one treats the charge operator as a spurion field \cite
{Manuel}, and assume it transforms conveniently under left/right
flavor symmetries, as well as under the color symmetries, depending
on whether the operator appears with flavor or color indexes, as it
will be shown below.

There are two important points that separate our treatment from
previous works. One is that we give the general form of the
low-energy Lagrangian in an arbitrary reference frame, that is, we
introduce all the possible covariant structures that can be formed
at finite density and in the presence of an external magnetic field.
When taken in the rest frame, the Lagrangian naturally reproduces
the different Lorentz structures that characterize the problem with
external magnetic field at finite density. The other is that when
proposing the allowed terms of the low energy Lagrangian, we take
into account that the coupling of the charged mesons with a rotated
electromagnetic field can be traced back to the coupling of the
fermions with the field. This coupling only occurs within a
covariant derivative to preserve the gauge invariance of the theory
under the $\widetilde U(1)$ group. Then, the coupling between the
Goldstone bosons and the rotated electromagnetic field always occurs
within a covariant derivative too. We will show that in this
$\widetilde U(1)$ gauge invariant approach, the charged Goldstone
bosons acquire field-induced masses that appear at the leading order
of the low-energy theory.

To find the effective low-energy theory, we can follow a similar
method to that used at zero field in
Refs. \cite{Casalbuoni:1999wu,Stephanov}. We start by introducing
two scalar fields which describe the fluctuations of the diquark
condensates, and are associated to left and right order
parameters:
\begin{equation}
X^{ai} \sim \epsilon^{abc} \epsilon^{ijk} \langle \psi^{bj}_L
\psi^{ck}_L \rangle^\dag \ , \qquad Y^{ai} \sim \epsilon^{abc}
\epsilon^{ijk} \langle \psi^{bj}_R \psi^{ck}_R \rangle^\dag
\end{equation}
where $i,j,k$ denote flavor indices, $a,b,c$ denote color indices,
and $L/R$ denote left/right chirality, respectively. Under an
$SU(3)_C \times SU(3)_L \times SU(3)_R$ rotation, the above fields
transform as
\begin{equation}\label{X}
X \rightarrow g_C X g_L^\intercal \ , \qquad Y \rightarrow g_C Y
g_R^\intercal \ ,
\end{equation}
with $g_C \in SU(3)_C $, and $g_{R/L} \in SU(3)_{R/L}$. The
expectation values of the $X$ and $Y$ fields define the ground state
of the CFL phase which produces the symmetry breaking $SU(3)_C
\times SU(3)_L \times SU(3)_R \rightarrow SU(3)_{C+L+R}$.

We are interested in the fluctuations of the phases of the order
parameters. Therefore, we can factor out the norm of the order
parameters in (\ref{X}) and work with unitary $X$ and $Y$ scalar
fields. Although the axial group $U(1)_{A}$ is anomalous at low
densities, it becomes an approximate symmetry at high densities
due to the suppression of the instanton interactions that produce
the anomaly. Nevertheless, in our derivations neither the pseudo
Goldstone mode associated to the breaking of this group, nor the
Goldstone mode associated to the $U(1)$-baryon symmetry breaking
will be considered, as these bosons are both neutral with respect
to the rotated electromagnetic charge, and therefore irrelevant
for the analysis of the present paper (we refer the interested
reader to Ref. \cite{Stephanov} to find out the contributions of
these modes to the low energy theory; see also Ref.
\cite{Ebert:2006tc} for the diquark excitations of the CFL ground
state at zero magnetic field in the framework of a
Nambu-Jona-Lasinio model).

We can introduce the Goldstone canonical fields
\begin{equation}\label{X&Y-Goldstones}
X=e^{i\Pi_{X}^{a}T_{a}}, \quad Y=e^{i\Pi_{Y}^{a}T_{a}}, \quad
a=1,...8
\end{equation}
with $T_{a}$ being the $SU(3)$ generators normalized to satisfy
\begin{equation}\label{Ta generators}
Tr[T_{a},T_{b}]=\frac{1}{2}\delta_{ab}
\end{equation}
Thus, X and Y together contain a total of sixteen scalar fields.
Strictly speaking only eight of them are genuine Goldstone bosons,
that is, massless scalar fields associated to the breaking of global
symmetries. The other eight are Higgs fields related to the
spontaneous breaking of the color gauge group SU(3) that gives mass
to the gluons and therefore can be always eliminated from the theory
by choosing a convenient (unitary) gauge \cite{Stephanov}. Since
these considerations remain valid in the presence of the external
magnetic field, we will work in the unitary gauge and concentrate
our analysis into the derivation of the low energy theory for the
genuine Goldstone bosons.

At zero magnetic field, the low energy theory of the Goldstone
bosons associated to the global symmetries can be written
\cite{Casalbuoni:1999wu} as
\begin{equation}\label{zeroB Lagrangian}
\mathcal{L}=-\frac{f_{\pi}^2}{4}Tr[(J^{\mu}_X-J^{\mu}_Y)^{2}]
\end{equation}
with currents $J^{\mu}_X$ and $J^{\mu}_Y$ defined by
\begin{equation}\label{zeroB currents}
J^{\mu}_X=X\partial^{\mu}X^{\dag}, \quad
J^{\mu}_Y=Y\partial^{\mu}Y^{\dag}.
\end{equation}
Notice that (\ref{zeroB Lagrangian}) does not take into account
the breaking of the Lorentz invariance due to the finite density.
The lack of Lorentz invariance was later incorporated in
Ref. \cite{Casalbuoni:1999wu} by manually assigning different
coefficients in front of the temporal and spatial derivative
terms.

Using (\ref{X}), we can verify that the currents $J_{X}$ and
$J_{Y}$ transform as
\begin{equation}\label{zeroB-currents-transformation}
J^{\mu}_{X}\rightarrow g_{C}J^{\mu}_{X}g_{C}^{\dag},\quad
J^{\mu}_{Y}\rightarrow g_{C}J^{\mu}_{Y}g_{C}^{\dag}
\end{equation}

The Lagrangian density (\ref{zeroB Lagrangian}) contains the
leading (second) order in derivatives terms invariant under the
$SU(3)_{C}\times SU(3)_{L}\times SU(3)_{R}$ rotations.

Two important changes occur when a $\widetilde{B}$ field is switch
on. First, the derivatives should be replaced by covariant
derivatives containing the rotated electromagnetic potential
$\widetilde{A}_{\mu}$ associated to the magnetic field
$\widetilde{B}$. Second, the number of fundamental tensors available
in the theory increases, because of the extra tensor
$\widetilde{F}_{\mu\nu}$. As a consequence, we can construct an
effective theory containing a larger number of independent terms
which are quadratic in the (covariant) derivative.

The covariant derivative should be consistent with the fact that
the VEV of the diquark fields, which only get contribution from
the diagonal elements of the order parameter matrix, are all
neutral in the rotated charge. We can generically denote the
$3\times3$ order parameter matrix by $\Delta$. Taking into account
that a rotated electromagnetic field is a combination of the $8^{th}$-gluon field and the conventional electromagnetic field and that
each element of the order parameter matrix carries an electric
charge equal to the sum of the electric charges of the quarks
forming the corresponding pair, we can write the covariant
derivative as
\begin{equation}\label{CD-1}
[\partial_{\mu}+igG_{\mu}^{8}(T^{8}\times
\mathbf{1})+ieA_{\mu}(\mathbf{1}\times
Q_{\Delta})]\Delta=[\partial_{\mu}-\frac{i\sqrt{3}}{2}gG_{\mu}^{8}(Q\times
\mathbf{1})-ieA_{\mu}(\mathbf{1}\times Q)]\Delta
\end{equation}
where the direct products denote (color $\times$ flavor).
$Q_{\Delta}$ is the conventional electromagnetic charge operator of
the quark pairs in the quark representation $(s,d,u)$. In this
representation the first column of $\Delta$ will have $(d,u)$ pairs,
the second $(s,u)$ pairs, and the third $(s,d)$ pairs. Hence
$Q_{\Delta}=diag(1/3,1/3,-2/3)$. Both the $T^{8}$ generator of the
$SU(3)$ group and the usual quark charge operator
$Q=diag(-1/3,-1/3,2/3)$ are connected to $Q_{\Delta}$ through the
relations $T^{8}=(\sqrt{3}/2)Q_{\Delta}=-(\sqrt{3}/2)Q$.

Taking into account that $\widetilde{A}_{\mu}=A_{\mu}cos \theta
-G_{\mu}^{8}sin \theta$, where $\theta$ is the mixing angle
\cite{alf-raj-wil-99/537}, Eq.(\ref{CD-1}) can be rewritten as
\begin{equation}\label{CD-2}
(\partial_{\mu}-i\widetilde{e}\widetilde{A}_{\mu}\widetilde{Q})\Delta
\end{equation}
where $\widetilde{e}=e\,  cos \theta=(\sqrt{3}/2)g\,  sin \theta$,
and $\widetilde{Q}=(\mathbf{1}\times Q- Q\times \mathbf{1})$ is
the "rotated" electric charge operator. As expected, the charge
operator $\widetilde{Q}$ assigns zero rotated charge to the
diagonal elements of $\Delta$. Notice that we do not keep the
orthogonal combination $\widetilde{G}_{\mu}^{8}=A_{\mu}sin \theta
+G_{\mu}^{8}cos \theta$, as the field $\widetilde{G}_{\mu}^{8}$
acquires a mass larger than $\Delta_{CFL}$, so it decouples from
the low-energy theory.

The straightforward generalization of the currents (\ref{zeroB
currents}) to the case with nonzero rotated magnetic field should be
done by substituting the derivative in (\ref{zeroB currents}) by the
covariant derivative (\ref{CD-2}). Taking into account that
$XX^{\dag}=1$ one can write the left current as
\begin{equation}\label{XY-currents}
(\widetilde{J}^{\mu}_X)_{ab}=X_{ai}(\partial^{\mu}\delta_{ij}+i\widetilde{e}\widetilde{A}^{\mu}
Q_{ij})X^{\dag}_{jb}-i\widetilde{e}\widetilde{A}^{\mu}Q_{ab}\nonumber
 \\
 =
 X(\partial^{\mu}X^{\dag}+i\widetilde{e}\widetilde{A}^{\mu}Q^{L}X^{\dag})
 -i\widetilde{e}\widetilde{A}^{\mu}Q^{C}
\end{equation}
where we introduced the notations $Q^{C}$ and $Q^{L}$ to keep
track of whether the operator $Q$ is an operator in color or
(left) flavor space. A similar expression can be found for the
right current $\widetilde{J}^{\mu}_Y$, with the obvious
substitution $X\rightarrow Y$ and $Q^{L}\rightarrow Q^{R}$.

As mentioned above, once a magnetic field is present one has an
extra tensor in the system that allows to create new structures in
the Lorentz space. Working in a covariant way, the leading order
low-energy Lagrangian density can be written as
\begin{equation}\label{L-1}
\mathcal{L}=-\frac{f_{\pi}^2}{4}Tr[(\widetilde{J}^{\mu}_X-\widetilde{J}^{\mu}_Y)
(\widetilde{J}^{\nu}_X-\widetilde{J}^{\nu}_Y)\Theta_{\mu\nu}]
\end{equation}
with
\begin{equation}\label{Structure}
\Theta_{\mu\nu}=C_{1} g_{\mu\nu}+C_{2}
u_{\mu}u_{\nu}+C_{3}\widehat{F}_{\mu\rho}\widehat{F}_{\rho\nu}
\end{equation}
being the most general Lorentz structure that can be formed with
the vectors and tensors available at low energies. In
(\ref{Structure}) we considered the normalized electromagnetic
tensor $\widehat{F}_{\mu \nu }=\frac{1}{ \left|
\widetilde{B}\right| }\widetilde{F}_{\mu \nu }$, in addition to
the usual metric tensor $g_{\mu\nu}$ and the vector four-velocity
of the center of mass of the dense medium $u_{\mu }$. The
$C_{i}'s$ are just constant coefficients. In this covariant
representation the magnetic field can be expressed as
$\widetilde{B}_{\mu }=\frac{1}{2}\varepsilon _{\mu \nu \rho
\lambda }u^{\nu }\widetilde{F}^{\rho \lambda }$. In the rest
frame, $u^{\nu }=(1,0,0,0)$, and the electromagnetic tensor
becomes $\widetilde{F}_{\mu\nu}=\widetilde{B}\delta_{\mu
1}\delta_{\nu 2}$ if we assume a magnetic field pointing along the
third spatial direction. In the absence of a magnetic field, the
coefficient $C_{3}$ is taken equal to zero and the Lagrangian
density (\ref{L-1}) reduces in the rest frame to that introduced
in Refs. \cite{Casalbuoni:1999wu,Stephanov}, where the system was
not Lorentz invariant due to the finite density ($u_{\mu}\neq 0$),
but it kept the rotational symmetry. In the presence of a magnetic
field the structure associated to $C_{3}$ naturally separates
between modes longitudinal and transverse to the field. A linear
term in $\widehat{F}_{\mu \nu }$ is forbidden, since it would
violate the theory CP invariance. Notice that in the leading
(second) order in derivatives we do not need to introduce the
momentum $k_{\mu}$ in the structures contributing to
(\ref{Structure}).

Notice that the flavor symmetry $SU(3)_{L}\times SU(3)_{R}$ is
explicitly broken by the electromagnetic coupling in (\ref{L-1}).
So, strictly speaking, (\ref{L-1}) is  invariant under
$\mathcal{G_{B}}$ but not under $\mathcal{G}$. However, one can
make the theory invariant under the original group $\mathcal{G}$
if we treat the charge operators as spurion fields and assume that
they transform as

\begin{equation}\label{Trans-currents}
Q^{L} \rightarrow g_{L}^{\ast}Q^{L}g_{L}^{\intercal},\qquad Q^{R}
\rightarrow g_{R}^{\ast}Q^{R}g_{R}^{\intercal},\qquad
Q^{C}\rightarrow g_{C}Q^{C}g_{C}^{\dag}.
\end{equation}
Using these transformations, one can show that the new currents
transform under $\mathcal{G}$ in the same way as the currents at
zero field:
\begin{equation}\label{B-currents-transformation}
\widetilde{J}^{\mu}_{X}\rightarrow
g_{C}\widetilde{J}^{\mu}_{X}g_{C}^{\dag},\quad
\widetilde{J}^{\mu}_{Y}\rightarrow
g_{C}\widetilde{J}^{\mu}_{Y}g_{C}^{\dag}
\end{equation}
yielding to the invariance of (\ref{L-1}) under $\mathcal{G}$.

 Following Ref. \cite{Casalbuoni:1999wu}, we introduce now the color singlet,
\begin{equation}
\Sigma = Y^{\dagger}X ,
\end{equation}
which transforms under $SU(3)_C \times SU(3)_L \times SU(3)_R$ as
$\Sigma \rightarrow g_R^{\ast} \Sigma g_L^{\top}$. In terms of
$\Sigma$ the Lagrangian density (\ref {L-1}) can be written as
\begin{equation}\label{L-2}
\mathcal{L}=\frac{f_{\pi}^2}{4}Tr[(D_{\mu}\Sigma
)(D_{\nu}\Sigma)^{\dag}\Theta_{\mu\nu}]
\end{equation}
where the covariant derivative acting on $\Sigma$ is

\begin{equation} \label{D-Sigma}
D_{\mu}\Sigma = \partial_{\mu}
\Sigma+i\widetilde{e}\widetilde{A}_{\mu} (Q^{R}\Sigma-\Sigma
Q^{L})
\end{equation}

In the rest-frame, (\ref{L-2}) becomes
\begin{equation}\label{chiral-lag}
{\cal L} = \frac{f_\pi^2}{4} \left[ {\rm Tr} \left(D_0 \Sigma)(
D_0 \Sigma^\dagger \right)+v_{\bot}^2 {\rm Tr} \left( D_{\bot}
\Sigma)( D_{\bot} \Sigma^\dagger \right) +v_{\|}^2 {\rm Tr} \left(
D_{\|} \Sigma )(D_{\|} \Sigma^\dagger \right) \right] \ .
\end{equation}
showing the expected separation between longitudinal $D_{\|}$ and
transverse $D_{\bot}$ (to the field) components of the covariant
derivatives, a direct consequence of the partial breaking of the
rotational symmetry in the presence of the magnetic field. The decay
constant $f_\pi$, and the meson maximum velocities $v_{\bot}$, and
$v_{\|}$, are parameters to be computed from the microscopic theory.
They can in general depend on the baryonic chemical potential and
the applied magnetic field. In the weak-field limit ($eB\ll\mu^2$),
at asymptotically large values of the chemical potential $\mu$, they
can be approximated by their zero-field values \cite{Stephanov}
\begin{equation}
\label{fpi} f_\pi^2 \approx {21 - 8 \ln{ 2} \over 18} \,
{\mu^2\over 2 \pi^2} \ , \qquad v_{\bot}\approx v_{\|} \approx
\frac{1}{\sqrt{3}}
\end{equation}

As we shall see in the next Section, the low-energy theory
(\ref{chiral-lag}) supports the generation of field-dependent masses
for the charged mesons at fields larger than a threshold value, even
though no quadratic-in-$\widetilde{B}$ term, like the one proposed
in Ref. \cite{Manuel} (Eq. (2.16)), is present at the leading order.
To understand this, one should keep in mind that the interaction
between $\Sigma$ and the electromagnetic field always occur through
the covariant derivative. To generate a quadratic-in-$\widetilde{B}$
(and quadratic-in-$Q$) term, one needs to consider a higher order
contribution involving four covariant derivatives
\begin{equation}
\label{higher-order-B2}\int d^{4}x Tr[(D_{\mu}\Sigma
)(D_{\nu}\Sigma)^{\dag}(D_{\rho}\Sigma
)(D_{\lambda}\Sigma)^{\dag}\widehat{F}_{\mu\nu}\widehat{F}_{\rho\lambda}]
\end{equation}
Using (we dropped total derivatives)
\begin{equation}
D_{\mu}\Sigma
(D_{\nu}\Sigma)^{\dag}\widehat{F}_{\mu\nu}=\frac{1}{2}\widehat{F}_{\mu\nu}
[i\widetilde{e}\widetilde{F}_{\mu\nu}(Q^{R}-\Sigma
Q^{L}\Sigma^{\dagger})
+2i\widetilde{e}\widetilde{A}_{\mu}(Q^{R}\Sigma\partial_{\nu}\Sigma^{\dagger}-\Sigma\partial_{\nu}\Sigma^{\dagger}
Q^{R})-2\Sigma(\partial_{\nu}\Sigma^{\dagger})\partial_{\mu}],
\end{equation}
it can be straightforwardly shown that the term proposed in
\cite{Manuel}
\begin{equation}
\label{higher-order-B2-Q2}\int d^{4}x
\widetilde{e}^{2}\widetilde{B}^{2}Tr[Q^{R}\Sigma Q^{L}\Sigma^{\dag}]
\end{equation}
is just one of the several terms coming out of
(\ref{higher-order-B2}). However, as proved below, the threshold
field for the decoupling of the charged mesons will be determined by
a contribution more relevant than (\ref{higher-order-B2-Q2}).

\section{CFL-MCFL Threshold Field}\label{threshold-field}

The unitary matrix $\Sigma$ can be parametrized in term of the
elementary Goldstone bosons $\phi^A$ as
\begin{equation}\label{Sigma}
 \Sigma =  \exp \left(i\,{\phi^A T^A \over f_\pi} \right) \ , \qquad
A = 1,...,8 \ ,
\end{equation}
where $T^A$ are the $SU(3)$ generators. Expanding (\ref{Sigma}) up
to linear terms in the fields, we can write (\ref{D-Sigma}) as a
sum of covariant derivatives for the $\phi^A$ fields
\begin{equation}\label{D-S-G}
 D_{\mu} \Sigma =  \frac{i}{f_\pi} [\sum_{A=1}^{3}
 T^{A}\partial_{\mu}\phi^{A}
 +T^{8}\partial_{\mu}\phi^{8}
 +\sum_{\pm}\tau^{\pm}(\partial_{\mu}\pm i\widetilde{e}\widetilde{A}_{\mu})\Pi^{{\pm}}
 +\sum_{\pm}\Lambda^{\pm}(\partial_{\mu}\pm i\widetilde{e}\widetilde{A}_{\mu})\kappa^{{\pm}}]
\end{equation}
where

\begin{equation}\label{+-}
\Pi^{{\pm}}=\frac{1}{\sqrt{2}}[\phi^{4}\mp i\phi^{5}], \qquad
\kappa^{{\pm}}=\frac{1}{\sqrt{2}}[\phi^{6}\mp i\phi^{7}],
\nonumber
\end{equation}

\begin{equation}
\tau^{{\pm}}=\frac{1}{\sqrt{2}}[T^{4}\pm iT^{5}], \qquad
\Lambda^{{\pm}}=\frac{1}{\sqrt{2}}[T^{6}\pm iT^{7}]
\end{equation}
and the rotated electromagnetic potential is taken in the Landau
gauge
\begin{equation}
\widetilde{A}_{\mu}=(0,0,\widetilde{B}x,0)
\end{equation}

From (\ref{chiral-lag}), the low-energy Lagrangian for the Goldstone
bosons can be written as

\begin{eqnarray}\label{chiral-lag-2}
L =\int d^{4}x \{
\frac{1}{4}[\sum_{A=1}^{3,8}|\partial_{0}\phi^{A}|^2+|\partial_{0}\Upsilon|^2]+\frac{v_{\|}^2}{4}[\sum_{A=1}^{3,8}|\partial_{\|}\phi^{A}|^2+|\partial_{\|}\Upsilon|^2]
\nonumber \\
+\frac{v_{\bot}^2}{4}[\sum_{A=1}^{3,8}|\partial_{\bot}\phi^{A}|^2+|(\partial_{\bot}+i\widetilde{e}\widetilde{A}_{\bot})\Upsilon|^2]\}
\qquad \qquad \qquad  \qquad \qquad
\end{eqnarray}
where we introduced the charged meson doublet

\begin{eqnarray}  \label{neutr-inv-propg}
\Upsilon \equiv \left(
\begin{array}{cc}
\Pi^{+}\\
\kappa^{+}
\end{array}
\right) \
\end{eqnarray}

The Lagrangian (\ref{chiral-lag-2}) represents the CFL low-energy
theory in the presence of a weak ($k^2 \sim {\tilde e}{\tilde
B}\ll\mu^2$) constant magnetic field. Now we are ready to
determine the strength of the threshold field for which the
effective symmetry transmutation from CFL to MCFL occurs.

For that aim, it is convenient to work in momentum space.
Transforming to momentum space in the presence of the magnetic field
can be done by applying to the scalar fields the same method
originally developed for fermions in \cite{Ritus:1978cj} and later
extended to vector fields  in \cite{efi-ext}. In this approach the
transformation to momentum space can be carried out using the wave
functions $S_{k}(x)$ of the asymptotic states of the charged mesons
in a uniform magnetic field. These functions play the role in the
magnetized medium of the usual plane-wave (Fourier) functions $e^{i
px}$ at zero field. Then, for the charged field $\Upsilon$ we have
\begin{equation}\label{P-Transformation}
\Upsilon (k)=\int d^{4} x S_{k} (x) \Upsilon (x)
\end{equation}
where
\begin{equation}\label{S}
S_{k}(x)=\mathcal{N} exp (ik_{0}x^{0}+ik_{2}x^{2}+ik_{3}x^{3})
D_{n}(\rho)
\end{equation}
with $D_{n}(\rho)$ being the parabolic cylinder functions with
argument $\varrho =\sqrt{2|\widetilde{e}\widetilde{B}|}(x_{1}- k_{2}/\widetilde{e}%
\widetilde{B})$, $\mathcal{N}$ the normalization constant
($\mathcal{N}=(4\pi| \widetilde{e}\widetilde{B}|)^{\frac{1}{4}}/
\sqrt{n!}$), and $n=0,1,...$ denoting the Landau levels. It is easy to check
that the transformation functions (\ref{S}) satisfy the
orthonormality condition

\begin{equation}\label{orthogonality}
\int_{-\infty}^{\infty} d^{4} x S_{k} (x) S_{k'} (x)=(2\pi)^{4}
\widehat{\delta}(k'-k)
\end{equation}
where $\widehat{\delta}(k'-k)=\delta  (k_0'-k_0)\delta
(k_2'-k_2)\delta(k_3'-k_3) \delta_{n'n} $.

Using this transformation in (\ref{chiral-lag-2}) we can derive
the Klein-Gordon equation for the charged mesons in momentum space

\begin{equation}\label{K-G}
[k_{0}^2 -\widetilde{e}\widetilde{B}(2n+1)v_{\bot}^2-k_{3}^2
v_{\|}^2]\Upsilon (k)=0.
\end{equation}
 and from it the corresponding dispersion relation
\begin{equation}\label{Dis-Eq}
E^2 = \widetilde{e}\widetilde{B}(2n+1)v_{\bot}^2+k_{3}^2 v_{\|}^2,
\end{equation}
We see that at zero momentum $(k_{3}=0, n=0)$ the rest energy of the
charged mesons is $M_{\widetilde{B}}^2 =
\widetilde{e}\widetilde{B}v_{\bot}^2$, meaning they acquire a mass
induced by the magnetic field. For a meson to be stable, its mass
should be less than twice the gap, otherwise it will decay into a
quasiparticle-quasihole pair. Here we are assuming that the
interactions between the Goldstone bosons and the quasiparticles are
described by a Yukawa term that originates within a NJL-type model.
Some authors (see \cite{quartic-quark model} for details) have
argued that the microscopic structure of the NG bosons in the CFL
phase should be that of a quartic-quark state. In the quartic-quark
picture the threshold would presumably be four times the gap energy.
Within the NJL-approach, the threshold field where the number of stable Goldstone bosons effectively changes, thus the symmetry of the CFL phase can not longer been treated as a good approximated symmetry is given by
\begin{equation}\label{Threshold-value}
\widetilde{e}\widetilde{B}_{MCFL}=\frac{4}{v_{\bot}^2}\Delta_{CFL}^2\simeq
12\Delta_{CFL}^2,
\end{equation}
where we used the weak-field approximation $v_{\bot}\simeq
1/\sqrt{3}$ \cite{Stephanov}. Contrary to the result found in
\cite{Manuel}, our threshold field does not depend on the decay
constant $f_{\pi}$, therefore it depends on $\mu$ only through
$\Delta_{CFL}$. The $f_{\pi}$ dependence found in \cite{Manuel} is a
direct consequence of considering a subleading contribution in the
derivation of the threshold field value, as previously shown.

For $\Delta_{CFL} \sim 15 MeV$ we get ${\tilde e} {\tilde B}_{MCFL}
\sim 10^{16}G$. At these field strengths, the charged mesons
decouple from the low-energy theory. When this decoupling occurs,
the five neutral Goldstone bosons (including the one associated to
the baryon symmetry breaking) that characterize the MCFL phase will
drive the low-energy physics of the system. Therefore, coming from
low to higher fields, the first magnetic phase that effectively
shows up in the magnetized system will be the MCFL, even though at
fields near the threshold field the splitting of the gaps found at
much stronger fields \cite{MCFL} may still be negligible.

We underline that the phenomenon occurring at the threshold field (\ref{Threshold-value}) is not a phase transition, as no symmetry is broken there. At any nonzero magnetic field strength, below or above the threshold field (\ref{Threshold-value}), the symmetry of the system is, strictly speaking, that of the MCFL. However, at $\widetilde{B}<\widetilde{B}_{MCFL}$ the charged Goldstone bosons are so light that the observable impact of the smaller symmetry of the MCFL phase, compared to that of CFL, is irrelevant and the nine Goldstone bosons of the CFL phase provide a good approximated description of the low-energy physics. Based on these considerations we choose to call the CFL-MCFL transition a symmetry transmutation.

On the other hand, it is worth to call attention to the analogy between the CFL-MCFL transmutation and what could be called a "field-induced" Mott transition. Mott transitions were originally considered in condensed matter in the context of metal-insulator transitions in strongly-correlated systems \cite{Mott-1}. Later on, Mott transitions have been also discussed in QCD to describe delocalization of bound states into their constituents at a temperature defined as the Mott temperature \cite{Mott-2}. By definition, the Mott temperature $T_{M}$ is the temperature at which the mass of the bound state equals the mass of its constituents, so the bound state becomes a resonance at $T>T_{M}$. In the present work, the role of the Mott temperature is played by the threshold field $\widetilde{B}_{MCFL}$. Mott transitions typically lead to the appearance of singularities at $T=T_{M}$ in a number of physically relevant observables. It is an open question worth to be investigated whether similar singularities are or not present in the CFL-MCFL transmutation at $\widetilde{B}_{MCFL}$.

\section{$\widetilde{B}$ \textit{vs} $\mu$ Phases in a Color
Superconductor with Three Quark Flavors}\label{mag-phases}

What will happen if we keep increasing the magnetic field until it
reaches the next energy scale $g\mu$? As known \cite{Vortex}, due to
the interaction of the applied magnetic field with the charged gluon
anomalous magnetic moment ($i\widetilde{e}\widetilde{f}_{\mu
\nu}G_{\mu}^{+}G_{\nu}^{-} $), once $\widetilde{B} \geq
\widetilde{B}_{PCFL} = m_{M}^2$, with $m_{M}\sim g\mu$ being the
magnetic mass of the charged gluons, one of the modes of the charged
gauge field becomes tachyonic (this is the well known "zero-mode
problem" for spin-1 charged fields in the presence of a magnetic
field found for Yang-Mills fields \cite{zero-mode}, for the
$W^{\pm}_{\mu}$ bosons in the electroweak theory \cite{Skalozub,
Olesen}, and even for higher-spin fields in the context of string
theory \cite{porrati}). Similarly to other spin-1 theories with
magnetic instabilities \cite{zero-mode}-\cite{Olesen}, the solution
of the zero-mode problem leads to the restructuring of the ground
state through the formation of an inhomogeneous gauge-field
condensate $G$, as well as an induced magnetic field due to the back
reaction of the $G$ condensate on the rotated electromagnetic field.
The magnitude of the $G$-condensate plays the role of the order
parameter for the phase transition occurring at $\widetilde{B}=
\widetilde{B}_{PCFL}$. Near the transition point, the amplitude of
the condensate $G$ is very small \cite{Vortex}. Then, the condensate
solution can be found using a Ginzburg-Landau (GL) approach similar
to Abrikosov's treatment of type II metal superconductivity near the
critical field $H_{c2}$ \cite{Abrikosov}. As in Abrikosov's case,
the order parameter $|G|$ continuously increases from zero with the
applied magnetic field, signalizing a second-order phase transition
towards a gluon crystalline vortex state characterized by the
formation of flux tubes. Both spatial symmetries -the rotational
symmetry in the plane perpendicular to the applied magnetic field
and the translational symmetry- are broken by the vortex state.

It turns out that, contrary to what occurs in conventional type-II
superconductors, where the applied magnetic field only penetrates
through flux tubes and with a smaller strength than that of the applied field, the gluon
vortex state exhibits a paramagnetic behavior. That is, outside the
flux tube the applied field $\widetilde{B}$ totally penetrates the
sample, while inside the tubes the magnetic field becomes larger
than $\widetilde{B}$. This antiscreening behavior is similar to that
found in the electroweak system at high magnetic field
\cite{Olesen}. Hence, since the $\widetilde{Q}$ photons remain
long-range in the presence of the condensate $G$, the
$\widetilde{U}(1)_{em}$ symmetry remains unbroken. At asymptotically
large densities, because $\Delta_{CFL}\ll m_{M}$, we have
$\widetilde{B}_{MCFL}\ll \widetilde{B}_{PCFL}$ for each $\mu$ value.

\begin{figure}
\begin{center}
\includegraphics[width=\textwidth]{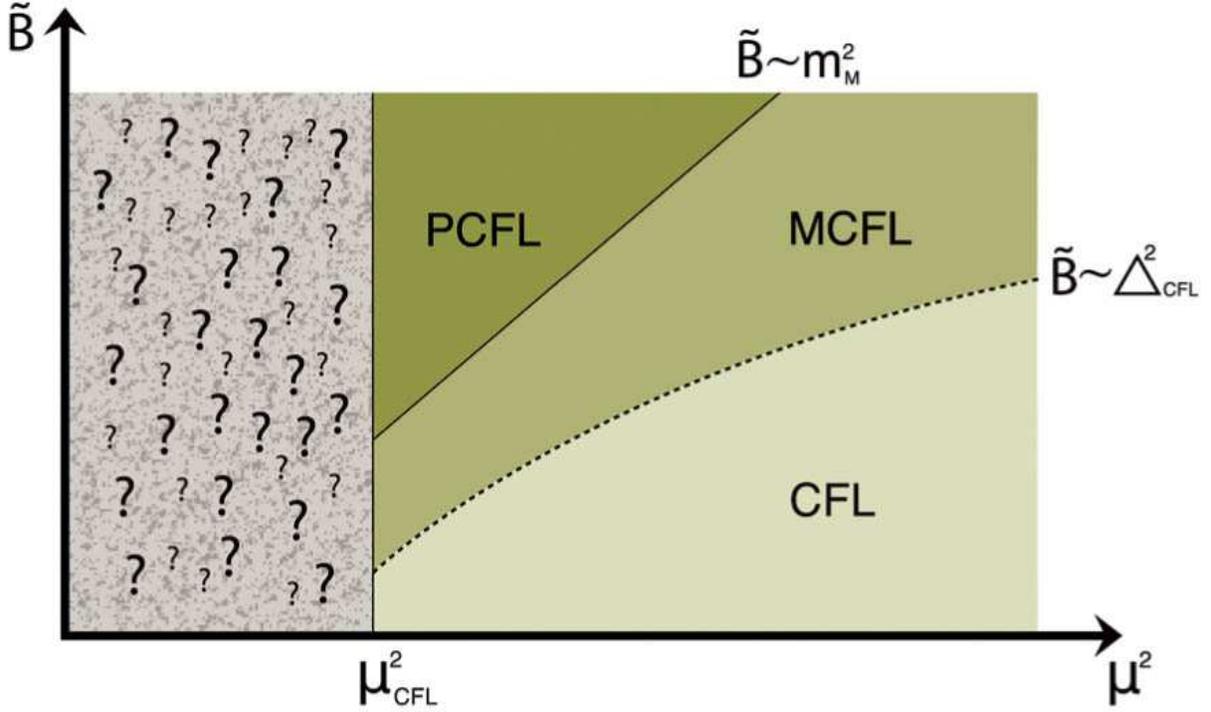}
\caption{Qualitative sketch in the $\widetilde{B}$ vs $\mu^2$ plane
of the different phases of a color superconductor with three quark
flavors in the presence of an external magnetic field at
asymptotically high densities.
The CFL phase appears here as an
approximate symmetry at weak field. Thus, the line between the CFL
and MCFL phases does not denote a real phase transition, but the
boundary separating the approximated CFL phase from the MCFL phase.
This symmetry-transmutation line is reached at field values of the
order of the CFL gap square. The line between the MCFL and PCFL
phases indicates a second-order phase transition curve occurring at
field strengths of the order of the magnetic mass square of the
charged gluons. The rectangular region to the left corresponds
to moderately high densities in the presence of a magnetic field. Since the
ground state at moderately high
densities has not yet been investigated in the presence of a magnetic field, this region is indicated by
question marks.} \label{fig1}
\end{center}
\end{figure}

At fields $\widetilde{B} \gtrsim \mu^{2}$ the density of states on
the Fermi surface of the charged quarks will be larger than that of
neutral quarks \cite{MCFL}. Because of the different density of states,
the magnitude of the gap receiving contributions
from pairs of rotated charged quarks will split from the magnitude
of the gap receiving contributions only from pairs of rotated
neutral quarks, as shown to happen, without taking into account the
vortex state, in Ref. \cite{MCFL}. The splitting of the gaps at this
scale, however, would not break any new symmetry that has not been
already broken at much lower scales. Hence, no new phase transition occurs at fields of order
$\mu^{2}$.

For fields much larger than $\mu^2$, i.e. sufficiently strong as to
surpass all the energy scales of the system, the quark infrared
dynamics will become predominant, and the phenomenon of magnetic
catalysis of chiral symmetry breaking \cite{MC} will be activated,
producing a phase that favors quark-antiquark condensates over
quark-quark condensates. However, the exploration of this region
goes beyond the scope of this paper.

Fig.\ref{fig1} provides a qualitative sketch of the magnetic phases
that exist at asymptotically high densities in the framework of a
three-flavor color superconductor. In that region, at very weak magnetic fields, the
color superconducting state is practically described by the CFL
phase, because the charged mesons, although massive, are so light
that they cannot decay in pairs of quasiparticle-quasihole. When the field
strength is of the order of the quarks' energy gap, the charged
mesons become heavy enough to decouple and the low-energy physics is
indeed that of the MCFL phase, where five neutral massless bosons
drive the low-energy behavior. At fields comparable to the magnetic
masses of the charged gluons, a chromomagnetic instability is
developed for these gluons leading to the formation of a vortex
state and the antiscreening of the magnetic field
\cite{Vortex,PCFL}. The vortex state breaks the translational
symmetry, as well as the remaining rotational symmetry in the plane
perpendicular to the applied magnetic field, hence the vortex
formation corresponds to a phase transition from the MCFL to a PCFL
phase.

The chemical potential $\mu_{CFL}$ in Fig.\ref{fig1} is used to
schematically separate the regions where the effects of an s-quark mass $M_{s}$
can (right of $\mu_{CFL}$) or cannot (left of $\mu_{CFL}$) be
neglected. The region of moderately high densities to the left of $\mu_{CFL}$
along the zero-magnetic-field line has been subjected to intense scrutiny in
the literature \cite{Alford}-\cite{Ferrer-Incera0705.2403}, as this is
the most important region for applications of CS in realistic systems
as compact stars. A main problem has been to find the stable phase at these moderately high densities. Despite many clever propositions that include a modified $CFL$-phase
with a condensate of kaons \cite{Schafer}; a LOFF phase
on which the quarks pair with nonzero total momentum \cite{LOFF};
as well as homogeneous \cite{miransky} and inhomogeneous \cite{Ferrer-Incera0705.2403} gluon condensate phases, it is still unclear which of these phases produces the lowest free energy. Given that nobody has yet studied this
question at finite magnetic field, we choose to indicate it in the figure with question marks.

\section{Concluding Remarks}\label{conclusions}
Summarizing, in a color superconductor with three-flavor quarks at
very high densities an increasing magnetic field produces a phase
transmutation from CFL to MCFL first, and then a phase transition
from MCFL to PCFL. During the phase transmutation no symmetry
breaking occurs, since in principle once a magnetic field is present
the symmetry is theoretically that of the MCFL, as discussed above.
However, in practice for $\widetilde{B}< \widetilde{B}_{MCFL}\sim
\Delta^{2}_{CFL}$ the MCFL phase is almost indistinguishable from
the CFL. Only at fields comparable to $\Delta^{2}_{CFL}$ the main
features of MCFL emerge through the low-energy behavior of the
system. At the threshold field $\widetilde{B}_{MCFL}$, only five
neutral Goldstone bosons remain out of the original nine
characterizing the low-energy behavior of the CFL phase. These are
precisely the five Goldstone bosons determining the new low-energy
behavior of the genuinely realized MCFL phase. Going from MCFL to
PCFL is, on the other hand, a real phase transition \cite{Vortex},
as the translational symmetry, as well as the remaining
rotational symmetry in the plane perpendicular to the applied
magnetic field are broken by the vortex state.

Throughout this paper we have ignored the effects due to quark
masses because we assumed very large baryon density. However, the
densities of interest for most astrophysical applications are just
moderately high. At moderate densities the s-quark mass can and do
play an important role \cite{Alford}. In this case, the color
superconductor develops chromomagnetic instabilities even in the
absence of an external magnetic field \cite{Fukushima}. At even
lower densities, where the s-quark decouples due to its larger mass,
a two-flavor color superconductivity -the so-called 2CS phase-
realizes. In this phase, when color neutrality and $\beta$
equilibrium conditions are imposed, some chromomagnetic
instabilities can also develop at certain density values
\cite{Igor}. Finding the stable superconducting ground state at
moderate densities is one of the main questions in the field at the
present moment \cite{Schafer}-\cite{Ferrer-Incera0705.2403}. The
understanding of this problem in the presence of a magnetic field,
which no doubt is another crucial player in the physical scenario of
a compact star, is also an important open question. In this regard,
as discussed in Ref. \cite{Ferrer-Incera0705.2403}, the removal of the chromomagnetic instabilities found at moderate
densities in the two-flavor system may be related to the
\textit{spontaneous generation} of an inhomogeneous gluon condensate
with a corresponding induced magnetic field. If this proposition is
proved to be correct also for the three-flavor case, the star's core could be in a PCFL-like
magnetic phase, even if the core's original magnetic field is zero
or relatively low. That is, the PCFL-like phase will not be
triggered by instabilities produced at a critical value of some
pre-existing inner magnetic field, but by instabilities connected to
the interplay of the neutrality conditions and the s-quark mass at
some baryon density. An interesting consequence of a PCFL core is
that a star's core in this phase can generate and/or boost its inner
magnetic field. Exploring the magnetic phases that realize at
realistic densities is an important pending task.

It is plausible that if compact stars are the natural playground for
color superconductivity, the magnetic phases described in this
paper, or more precisely, the version of these phases at more
realistic densities, may be relevant for the physics of the core of
highly magnetized compact objects like magnetars
\cite{magnetars,transient magnetar}, and the so-called Central
Compact Objects (CCO) \cite{CCO}. CCO are point-like sources located
near the center of supernova remnants that cannot be identified as
active radio pulsars or magnetars \cite{CCO}. Some of them may have
magnetar's strength B fields and much smaller radii. For typical
neutron star masses $\sim 1.4 M_{\odot}$, a smaller radii means a
denser star's core; the denser the core, the greater the chance it
can be in a color superconducting phase. Magnetars' surface magnetic
fields are typically in the range of $10^{14}G-10^{16}G$, and the
fields at their much denser cores can probably reach even larger
values. Any of these compact objects could be a candidate for the
realization of a color superconducting phase in a highly magnetized
background.

Moreover, the standard explanation of the origin of the magnetars'
large magnetic fields cannot explain all the features of the
supernova remnants surrounding these objects
\cite{magnetar-criticism,Xu06}. Magnetars are supposed to be created
by a magnetohydrodynamic-dynamo mechanism that amplifies a seed
magnetic field due to a rapid rotating (spin period $<3 ms$)
protoneutron star. Part of this rotating energy is supposed to power
the supernova through rapid magnetic braking, implying that the
supernova remnants associated with magnetars should be an order of
magnitude more energetic than typical supernova remnants. However
recent calculations \cite{magnetar-criticism} indicate that their
energies are similar. In addition, one would expect that when a
magnetar spins down, the rotational energy output should go into a
magnetized particle wind of ultrarelativistic electrons and
positrons that radiate via synchrotron emission across the
electromagnetic spectrum. Nevertheless, so far nobody has detected
the expected luminous pulsar wind nebulae around magnetars
\cite{Safi-Harb}.

Although more observations are needed to confirm the above, current
observations indicate that alternative models to the standard
magnetar model \cite{magnetars} need to be considered. For example,
some authors \cite{magnetar-criticism} have suggested that magnetars
could be the outcome of a stellar progenitor with highly magnetized
cores. A progenitor star with a PCFL-like core would be capable to
induce and/or enhance the star's magnetic field due to the antiscreening mechanism
inherent to this color superconducting phase \cite{Vortex,PCFL}, and
as such provides an alternative to the observational conundrum of
the standard magnetar's paradigm \cite{magnetars}.

Only after the stable color-superconducting phase at moderate
densities is well established with and without an external magnetic
field, we will be in a well-grounded position to investigate and
reliably predict observable signatures of color superconductivity in
compact stars.

{\bf Acknowledgments:} We are grateful to C. Manuel for comments and
to S. Safi-Harb and I.~A.~ Shovkovy for insightful discussions and
comments. We also like to thank the referee for helpful comments and for calling
our attention to the analogy with a Mott transition.
This work has been supported in part by DOE Nuclear Theory
grant DE-FG02-07ER41458.

\end{document}